# Detecting overlapping community structure: Estonian network of payments

Stephanie Rendón de la Torre[*], Jaan Kalda[*], Robert Kitt[*1], Jüri Engelbrecht[*]

[*]School of Science, Department of Cybernetics, Tallinn University of Technology, Akadeemia tee 21, 12618, Tallinn, ESTONIA
[1]Swedbank AS, Liivalaia 12, 15038, Tallinn, ESTONIA

________________________________________________________________________

Revealing the community structure exhibited by real networks is a fundamental phase towards a comprehensive understanding of complex systems beyond the local organization of their components. Community detection techniques help on providing insights into understanding the local organization of the components of networks. In this study we identify and investigate the overlapping community structure of an interesting and unique case of study: the Estonian network of payments. In order to perform the study, we use the Clique Percolation Method and explore statistical distribution functions of the communities, where in most cases we found scale-free properties. In this network the nodes represent Estonian companies and the links represent payments done between the companies. Our study adds to the literature of complex networks by presenting the first overlapping community detection analysis of a country's network of payments.

*Keywords:* Complex networks; economic networks; overlapping communities; scale-free networks

## Kattuvate kommuunide tuvastamine Eesti maksevõrgustikus

Reaalselt eksisteerivates võrgustikes sisalduvate kommuunide tuvastamine on üheks põhietapiks teel komplekssüsteemide selliste seaduspärasuste mõistmise poole, mis lähevad sügavamale üksikelementide lokaalsete interaktsioonide käsitlemisest. Kommuunide tuvastamise meetodid aitavad heita valgust võrgustike komponentide lokaalstruktuuridele. Käesolevas uurimuses identifitseerime ja uurime kattuvate kommuunide struktuure olulises unikaalses võrgustikus - Eesti maksetevõrgustikus. Selleks otstarbeks kasutame nn klikk-perkolatsiooni meetodit ja uurime kommuunide jaotusfunktsioone ning kommuunide mastaabi-invariantseid omadusi. Antud võrgustikus on sõlmpunktideks Eesti ettevõtted ja sidemeteks maksed erinevate ettevõtete vahel. Tegemist on esmakordse uurimusega, kus tuvastatakse kattuvad kommuunid ühe riigi ettevõtete vaheliste maksete võrgustikus

# 1. Introduction

A network is a set of nodes connected by links. A complex network has nontrivial topological features and most of the real-world networks are complex. Complex networks can be described by a combination of local, global and mesoscale approaches. The exploration of intermediate-sized structures that are responsible for "coupling" local properties demands partitioning networks into useful groups of nodes [1]. Networks have sections in which the nodes are more densely connected to each other than to the rest of the nodes in the network, and such sub-sections are called communities. Communities might exist in networked systems of different nature, such as: economics, sociology, biology, engineering, politics and computer science.

Community detection is a graph partitioning process that provides valuable insight of the organizational principles of networks and is essential for exploring and predicting connections that are not yet observed. Thus far, recent advances of the underlying mechanisms that rule dynamics of communities in networks are limited, and this is why the achievement of an extensive and wider understanding of communities is important. Locating the underlying community structure in a network allows studying the network more easily and could provide insights into the function of the system represented by the network, as communities often correspond to functional units of systems. The study of communities and their properties also helps on revealing relevant groups of nodes, creating meaningful classifications, discovering similarities or revealing unknown linkages between nodes. Communities have a strong impact in the behavior of a network as a whole and studying them is fundamental in order to expand the knowledge of the community structure beyond the local organization of the components of networks.

The usefulness of identifying the communities within networks lies in how this information could be used in a practical scenario. Particularly, in the context of the bank industry the output of our community analysis (based on payments between companies which are customers of a bank) could be used for targeted marketing. For example, it could be used at the moment of integrating criteria for creating target groups of customers to whom certain products or lines of products would be offered. Customers in the same community would be included in the same target group and later on after one offer is made to them it would be possible and interesting to assess the contagion effects of the product acquisition among customers of the same communities who received the same offer. Another useful application is when helping to create customer-level segmentations or marketing profiles. To know the community (or communities) where a customer belongs to, could be one of the main features for creating customer profiles or clustering levels. An alternative usage of the output of community analysis is in predictive analytics for example when building churn models. Churn models usually define a measure of the potential risk of a customer cancelling a product or service and provide awareness and metrics to execute retention efforts against churning. The communities to which the companies/customers belong to could be used as variables or features when using logistic regression, random forest or neural network models. Additionally, community detection analysis could be used as input for product affinity and recommender systems. Affinity analysis is a data mining technique that helps to group customers based on historical data of purchased products and is used for cross-selling product recommendations. Another useful and immediate application is in product acquisition propensity models. These models calculate customers' likelihood to acquire a product after an offer is made based on a myriad of variables and with this evidence the sales process can become more efficient.

The objective of this study is to detect the overlapping community structure of the large-scale payments network of Estonia by examining its characteristics and scale-free properties through the Clique Percolation Method [2-3]. First, we detect communities and then we analyze the global structure of the network through the distribution functions of four basic quantities.

The research questions for this study are the following: Which is the community structure of the Estonian network of payments? Are there scale-free properties in the community structure?

Section 1 provides a general introduction and an overview of the objectives. In Section 2 we deliver a description of the data set used in this study. Section 3 provides a literature review of studies related with similar networks and their applications. In Section 4 we present the method used to develop this study, while Section 5 presents our main results and findings. Finally, Section 6 concludes with a discussion of our results.

**2. Analyzed data**

Our data set was obtained from Swedbank's databases. Swedbank is one of the leading banks in the Nordic and Baltic regions of Europe. The bank operates actively in Estonia, Latvia, Lithuania and Sweden. All the information related to the identities of the nodes is very sensitive and thus will remain confidential and unfortunately cannot be disclosed. The data set is unique in its kind and very interesting since ~80% of Estonia's bank transactions are executed through Swedbank's system of payments, hence, this data set reproduces well the transactional trends of the whole Estonian economy, so we use this data set as a proxy of the Estonian economy.

The data set consists of electronic company-to-company domestic payments, including data of 16,613 companies and 3.4 million payment transactions (October 2013-December 2014). In this study, the nodes represent companies and the links represent the payments between the companies. For simplicity, we focus on the basic case where the network of payments is defined by a symmetric payment adjacency matrix that represents the whole image of the network. We considered an undirected graph approach where two nodes have a link if they share one or more payments. Then each element represents a link if there is a transaction between company $i$ and $j$, as follows: $a_{ij}^u = a_{ji}^u$ where $a_{ij}^u = 1$, $a_{ij}^u = 0$ if there is no transaction between $i$ and $j$.

Tables 1 and 2 show main measures and statistics of our network of payments. The average degree of our network is $\langle k \rangle = 21$ while the diameter is 29. The average betweenness of links is 41 while is 112 for nodes. The average shortest path length $\langle l \rangle = 7.3$. Our network is a "small world" with 7 degrees of separation, so in average any company can be reached by another within seven steps. An average degree of separation of 7 is a very small value for a network of size $N = 16,61$. The network displays scale-free properties in the degree distribution. The degree distribution follows a power-law where the scaling exponent is: $P(\geq k) \propto k^{\wedge}(-2.46)$. The network has a low average clustering coefficient of 0.19 and displays disassortative mixing behavior, where high degree nodes, represented by companies who have many counterparties such as business partners, service providers, clients or suppliers, have a large number of links to companies which have only one link, or just few links.

Table 1 Network's characteristics

| Number of companies analyzed | 16,613 |
|---|---|
| Total number of payments analyzed | 3,406,651 |
| Total value of transactions | 4,342,109,265 * |
| Average value of transaction per customer | 99,904 * |
| Maximum value of a transaction | 135,736 * |
| Minimum value of a transaction (aggregated) | 1,000 * |
| Average volume of transaction per company | 76 |
| Maximum volume of transaction per company | 34,665 |
| Minimum volume of transaction per company (aggregated) | 20 |

*All monetary quantities are expressed in monetary units and not in real currencies in order to protect the confidentiality of the data set. The purpose of showing monetary units is to provide a notion of the proportions of quantities and not to show exact amounts of money.

Table 2 Summary of Statistics

| Statistic | Value |
|---|---|
| $<k>$ | 21 |
| $\gamma^o$ | 2.41 |
| $\gamma^i$ | 2.50 |
| $\gamma$ | 2.46 |
| $<C>$ | 0.19 |
| $<l>$ | 7.3 |
| $T$ | 0.13 |
| $D$ | 29 |
| $<\sigma>$ (nodes) | 112 |
| $<\sigma>$ (links) | 41 |

N = number of nodes. $<k>$ = average degree. $\gamma^o$ = scaling exponent of the out-degree distribution. $\gamma^i$ = scaling exponent of the in-degree distribution. $\gamma$ = scaling exponent of the connectivity degree distribution. $<C>$ = average clustering coefficient. $<l>$ = average shortest path length. $T$ = connectivity % . $D$ = Diameter. $<\sigma>$ = average betweenness. More information on this network's statistics can be found in [40].

## 3. Literature discussion

Networks play an important role in a wide range of economic and social phenomena and the use of techniques and methods from graph theory has permitted economic network theory to expand the knowledge and understanding of economic phenomena in which the embeddedness of individuals or agents in their social or economic interrelations cannot be ignored [4]. For example, Souma et *al*. [5] studied a shareholder network of Japanese companies where the authors analyzed the companies' growth through economic networks dynamics. Other examples of interesting applications of complex networks in economics are provided by the regional investment or ownership networks where European company-to-company investment stocks show power-law distributions that allow predicting the investments that will be received or made in specific regions, based on the connectivity and transactional activity of the companies [6-7]. Nakano and White [8] have shown that analytic concepts and methods related with complex networks can help to uncover structural factors that may influence the price formation for empirical market-link formations of economic agents. Reyes et *al.* [9] used a weighted network analysis focused on using random walk betweenness centrality to study why high-performing Asian economies have higher economic growth than Latin-American economies in the last years. Network-based approaches are very useful and provide a means by which to monitor complex economic systems and may help on providing better control in managing and governing these systems. Another interesting line of research is related with network topology as a basis for investigating money flows of customer driven banking transactions. A few recent papers describe the actual topologies observed in different financial systems [10-13]. Other works have focused on economic shocks and robustness in economic complex networks [14-15].

Regarding community studies on economic networks and their applications, Vitali and Battiston [16] studied the community structure of a global corporate network and found that geography is the major driver of organization within that network. Also, in this study they assessed the role of the financial sector in the architecture of the global corporate network by analyzing centralities of communities. Fenn et *al.* [17] studied the evolution of communities of a foreign exchange market network in which each node represents an exchange rate and each link represents a time-dependent correlation between the rates. By using community detection, they were able to uncover major trading changes that occurred in the market during the credit crisis of 2008. Other economic communities' studies have focused on the overlapping feature of communities, such as in [18-19].

General community detection studies on other types of networks have been devoted to study communities representing real social groupings [20-22], communities in a co-authorship network representing related publications of specific topics [23], protein-protein interaction networks [24], communities in a metabolic network representing cycles and functional units in biology [25-26] and communities in the World Wide Web representing web pages with related contents [27].

Most algorithms for community detection can be distinguished in divisive, agglomerative and optimization-based methods and each one has specific strengths and weaknesses. Previous studies on network communities based on divisive and agglomerative methods consider that structures of communities can be expressed in terms of separated groups of clusters [28-31], but most of the real networks are characterized by well-defined statistics of overlapping communities. An important limitation of the popular node partitioning methods is that a node must be in one single community whereas it is often more appropriate to attribute a node to several different communities, particularly in real-world networks. An example where community overlapping is commonly observed is in social networks where individuals typically belong to many communities such as: work teams, religious groups, friendship groups, hobby clubs, family or other similar social communities. Moreover, members of social communities have their own communities and this in turn results in a very complex web of communities [3]. The phenomenon of community overlapping has been already noticed by sociologists but has been barely studied systematically for large-scale networks [2,32-35].

## 4. Method

Overlapping communities arise when a node is a member of more than one community. In economic systems the nodes could frequently belong to multiple communities, therefore, forcing each node to belong into a single community might result into a misleading characterization of the underlying community structure. The Clique Percolation Method (CPM) [2-3] is based on the assumption that a community comprises overlapping sets of fully connected subgraphs and detects communities by searching for adjacent cliques. A clique is a complete (fully connected) subgraph. A $k$-clique is a complete subgraph of size $k$ (the number of nodes in the subgraph). Two nodes are connected if the $k$-cliques that represent them share $k - 1$ members. The method begins by identifying all cliques of size $k$ in a network. When all the cliques are identified, then a $N_c \times N_c$ clique-clique overlapping symmetric matrix $\boldsymbol{O}$ can be constructed, where $N_c$ is the number of cliques and $\boldsymbol{O}_{ij}$ is the number of nodes shared by cliques $i$ and $j$ [36]. This overlapping matrix $\mathbf{O}$ encodes all the important information needed to extract the $k$-clique communities for any value of $k$. In the overlapping matrix $\mathbf{O}$, rows and columns represent cliques and the elements are the number of shared nodes between the corresponding two cliques. Diagonal elements represent the size of the clique and when two cliques intersect they form a community.

For certain $k$ values, the $k$-clique communities form such connected clique components in which their nearby cliques are linked to each other by at least $k-1$ adjacent nodes. In order to find these components in the overlapping matrix $\mathbf{O}$, one should keep the entries of the overlapping matrix which are larger than or equal to $k-1$, set the others to zero and finally locate the connected components of the overlapping matrix $\mathbf{O}$. Communities correspond to each one of the identified separated components [2].

## 5. Results

For CPM is important to choose a parameter $k$. The parameter $k$ affects the constituents of the overlapping regions between communities. The larger the parameter $k$, the less the number of nodes which can arise in the overlapping regions. When $k \to \infty$, the maximal clique network is identical to the original network and no overlap is identified. The choice of $k$ will depend on the network. It is observed from many real-world networks, that the typical value of $k$ is often between 3 and 6 [37].

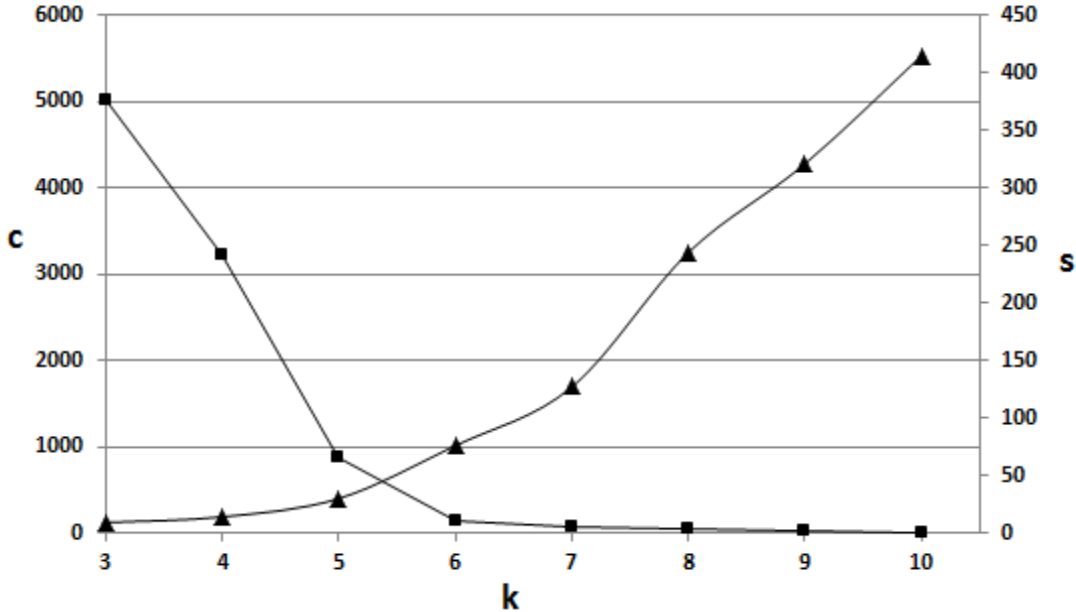

Fig. 1 Plot of the average size of community (s) and number of communities (c) as $k$ increases. Squares represent the number of communities and triangles represent the size of the communities.

Fig. 1 shows a plot of the number of communities and the average size of the communities at different $k$ values. As $k$ increases the number of communities decreases while the size of the communities increases rapidly. When $k$ decreases the number of communities increases rapidly while the size of the communities remains low. In order to obtain the optimal value of $k$, we tested different values ranging from 3 to 10 and *a posteriori* we chose $k = 5$ because when $k < 5$ a high number of communities arises and the partitions become very low and giant communities appear (with sizes of more than 3200); at the level $k = 5$ we obtain a rich partition with the most widely distributed cluster sizes set for which no giant community appears.

An overlapping community graph is a representation of a network that denotes links between communities, where the nodes represent the communities and the links are represented by the shared nodes between communities. For visualization purposes and in order to draw a readable map of the network, Fig. 2 shows a graphic view of a representative section of the overlapping network of communities where big and small communities can easily be distinguished. Fig. 2 depicts 25 overlapping communities and each colored circle represents a node which in turn represents an overlapping community. The links represent the shared nodes between the communities. The size of the nodes characterizes the size of each community. For example, the big node in the middle represents a community with 61 companies.

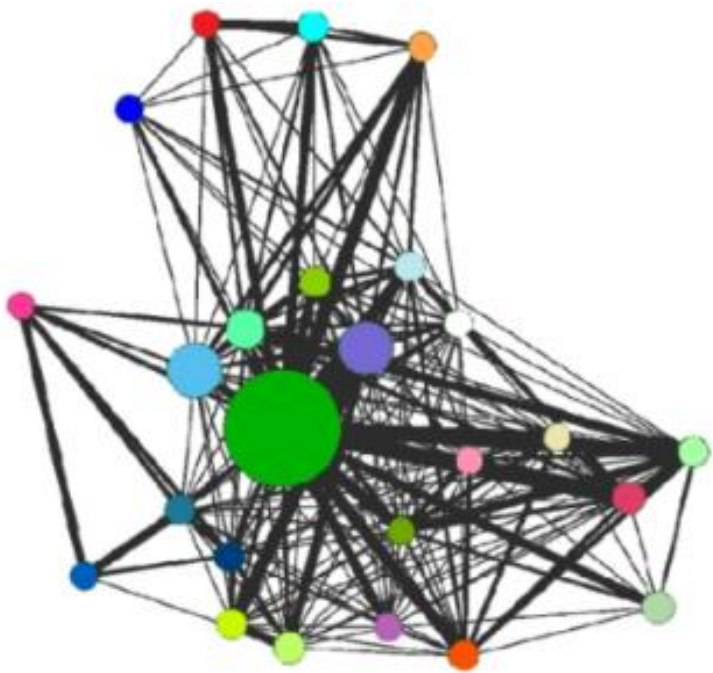

Fig. 2 Visual representation of a section of the overlapping network of communities (Estonian network of payments). The circles (nodes) represent communities and the black lines between them represent shared nodes between communities.

### 5.1. Structure of communities

In order to study and characterize the global community structure of our network, we investigated the distribution functions of the following four elementary quantities: community size $P(s)$, overlap size $P(s_o)$, community degree $P(d)$ and membership number $P(m)$. The aforementioned distributions are shown in Figs. 3-7. In general, nodes in a network can be characterized by a membership number which is the number of communities a node belongs to. This means that for example, any two communities may share some of their nodes which correspond to the overlap size between those communities. There is also a network of communities where the overlaps are represented by the links and the communities are represented by the nodes, and the number of such links is called: community degree. The size of any of those communities is defined by the number of nodes it has.

The community size distribution is an important statistic that describes partially the system of communities. Fig. 3 displays the cumulative distribution function of the community size $P(s)$ and it shows the probability of a community to have a size higher or equal to $s$ calculated over different points in time, where $t$ is the time in months. The overall distribution of community sizes resembles a power-law $P(s) \propto s^\alpha$, where $\alpha$ is the scaling exponent, and a power-law is valid nearly over all times $t$. The scaling exponent (calculated by maximum likelihood estimators) when $t = 3$ is -2.8 (included for eye guideline) and Eq. 1 is:

$$P(s) \propto s^{-2.8}. \tag{1}$$

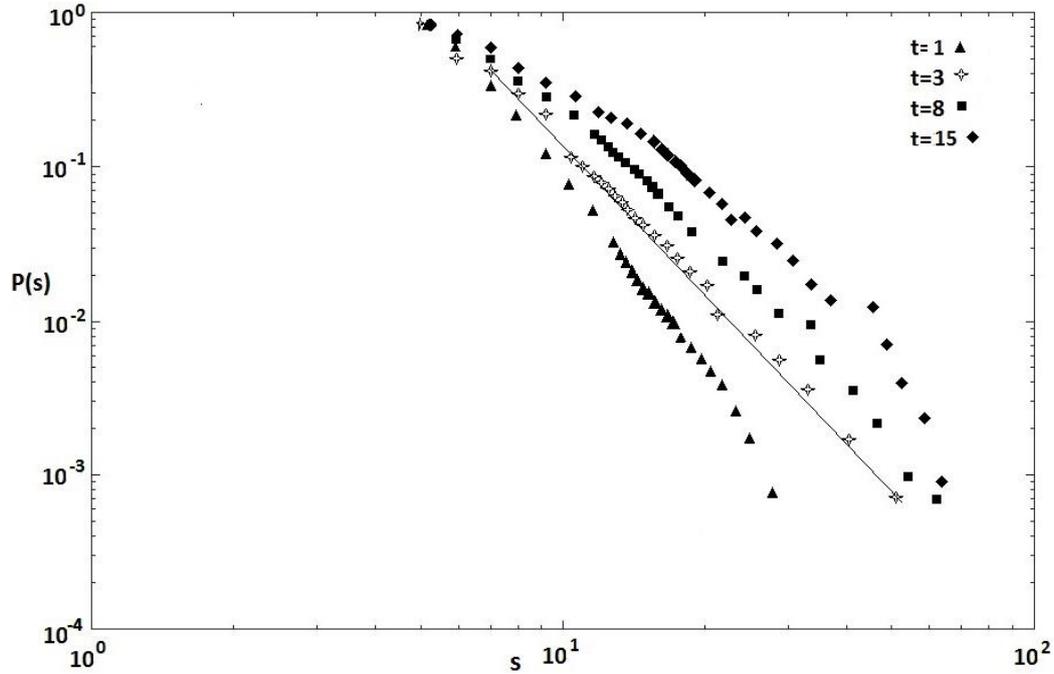

Fig. 3 Cumulative community size distribution at different times $t$ (log-log scale)

The sizes of the communities at $t = 1$ are smaller than in the rest of the months; as time increases the size increases, particularly the size of the largest communities. The shapes of the power-laws observed in the community size distributions of Fig. 3 suggest there is no characteristic community size in the network. The distribution at different moments in time follows similar decaying patterns, but in general, the scaling tail is higher as $t$ increases. A fat tail distribution implies that there are numerous small communities coexisting with few large communities [38-39]. Fig. 4 shows statistics of the community sizes across time and according to the plot, both the standard deviation of community sizes and the average size of communities increased with time.

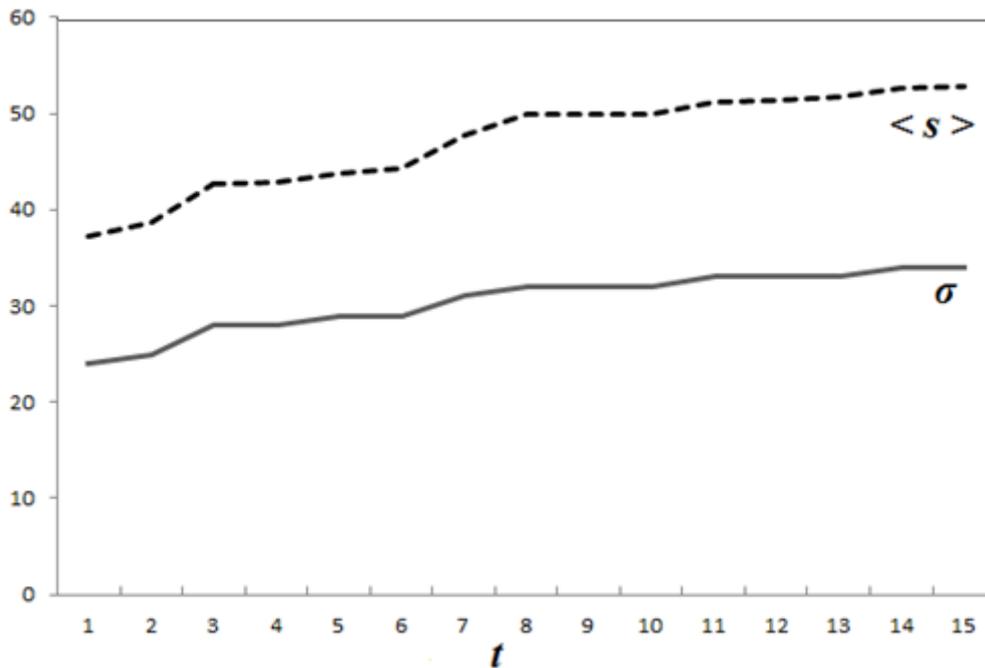

Fig. 4 Statistics of community size. <s> is the average community size. σ is the standard deviation of the size of communities at different times *t*.

In a network of overlapping communities, the overlaps are represented by the links and the number of those links is represented by the community degree $d$. Then, the degree $d$ is the number of communities another community overlaps with. Fig. 5 shows the cumulative distribution of the community degrees in the network. There are some outstanding community degrees by the end of the tail and these include communities that cluster the majority of the biggest customers in the network. The central part of the distribution decays faster than the rest of the distribution. There is an observable curvature in the log-log plot, however no approximation method fitted the distribution. Fig. 5 shows that the maximum number of degrees $d$ is 63 and corresponds to a relatively small quantity of nodes.

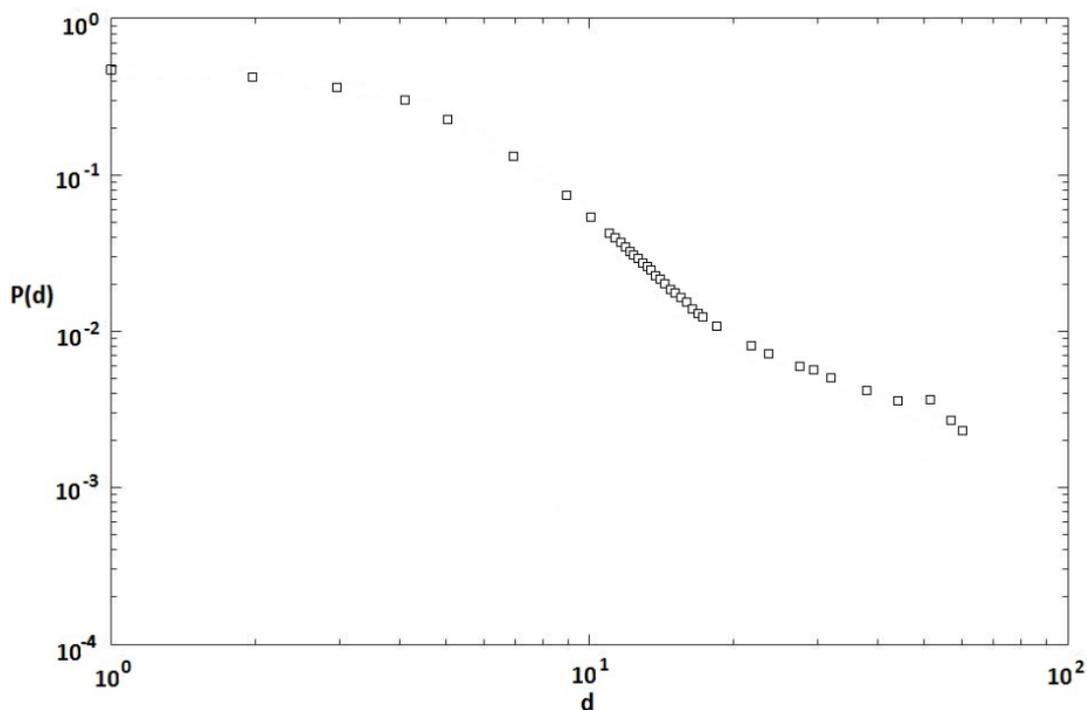

Fig. 5 Cumulative distribution of community degrees $d$ (log-log scale)

A node $i$ of a network can be characterized by a membership number $m_i$, which is the number of communities where the node $i$ belongs to. Fig. 6 shows the cumulative distribution of the membership number $m_i$. The distribution follows a power-law where no characteristic scale exists. The largest membership number found in the network was 10, meaning that a company can belong to maximum 10 different communities simultaneously. Fig. 6 shows that the fraction of nodes that belong to many different communities is quite small, while the fraction of nodes belonging to at least 1 community is high. For example, when $m = 1$ the percentage of nodes that belong to at least one community is 50%, while the percentage of nodes that belong simultaneously to 10 communities ($m = 10$) is extremely small. However the rest of the communities belong to at least 2 or more communities. The companies that overlap with 10 communities belong to the energy and water services. The majority of the nodes that have $m \neq 1$ have a degree that is less than $k - 1$, meaning they are weakly connected.

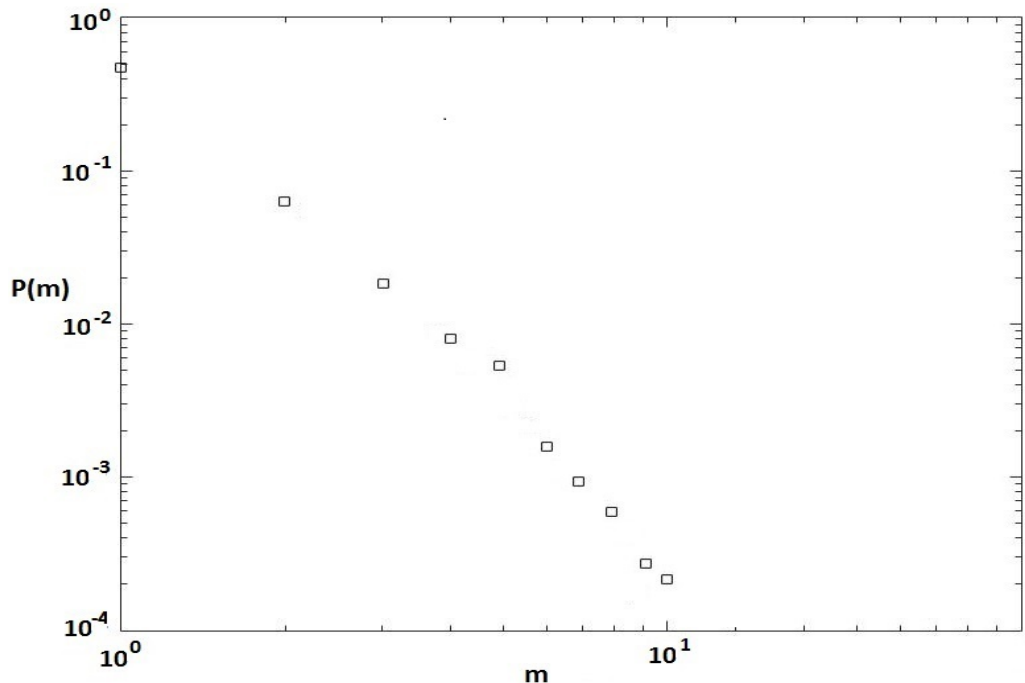

Fig. 6 Cumulative distribution function of the membership number $m_i$. (log-log scale)

The range to which the communities overlap with each other is also an important property of our network. The overlap size is defined as the number of nodes that two communities share. $P(s_o)$ is the proportion of overlaps larger than $s_o$. Fig. 7 shows the cumulative distribution function of the overlap size. In general, although the extent of overlap sizes is limited, the data is close to power-law dependence, meaning there is no characteristic overlap size. The largest overlap size is 22, however at $s_o \geq 9$ the number of overlapping nodes becomes small.

In our previous study [40] we found scale-free properties in the degree distributions of the Estonian network of payments and it is interesting to observe that the scale-free property is also preserved at a higher level of organization where overlapping communities are present. Scale-free networks are resilient against random removal of nodes and this means that is difficult to destroy a complex network by random mechanisms, but if the exact portion of particularly selected nodes are removed, then the network breaks easily. When the degree distributions of networks present scale-free structure, then this fact determines the topology of the system. Scale-free networks are robust against random damages but they are vulnerable against targeted attacks of nodes,

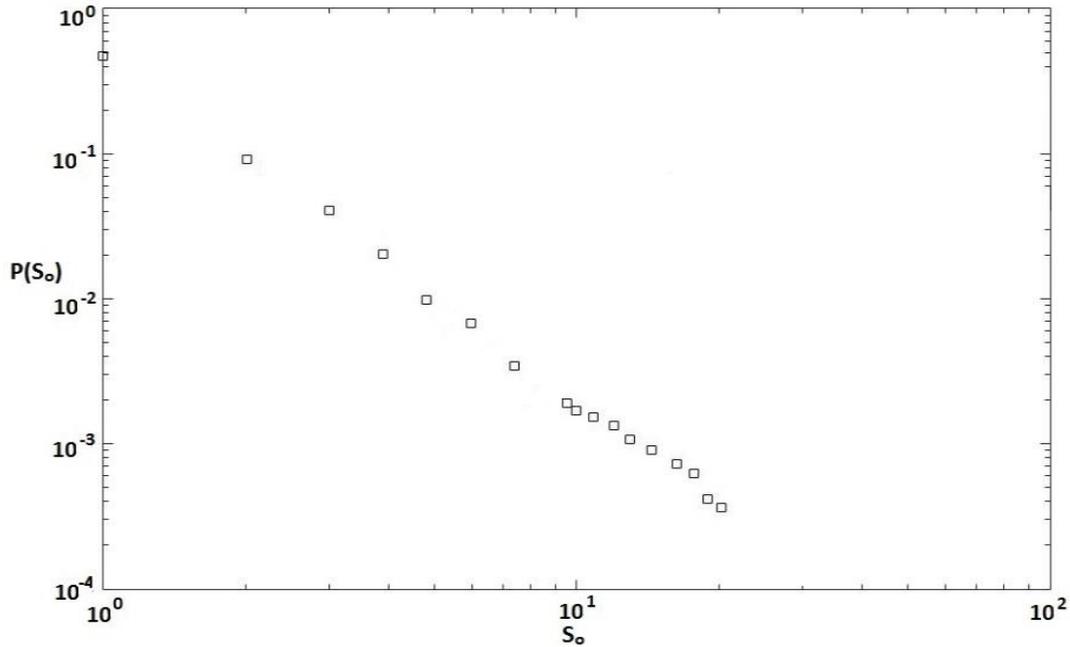

Fig. 7 Cumulative distribution function of the overlap size $s_o$. (log-log scale)

## 6. Conclusions

In this study we have analyzed the community structure of the Estonian network of payments by using the Clique Percolation Method. We found that there are scale-free properties in the statistical distributions of the community structure. Size, overlap and membership distributions follow shapes that are compatible with power-laws. Power-law distributions have already appeared in this network at a global scale in the level of nodes [40], and in this community structure study we have shown that power-laws are present at the level of overlapping communities as well. This study adds to the existing literature of complex networks by presenting the first overlapping community analysis of a country's network of payments.

An immediate application and usefulness for the community detection output is that it can be used in targeted marketing activities, as input for predictive analytical models such as product acquisition propensities, churn propensities, product affinity analyses, for creating marketing profiles or customer segmentations and for creating customer target lists for product offering (in an effort to propagate consumer buzz effects). Further applications for community detection in similar economic networks could involve strengthening relationships between companies of the same community for improving performance of the whole network, identification of patterns between companies and tracking suspicious business activities.

A question that remains open for future research is to investigate if the similarities in communities' features amongst different complex networks arise randomly or if there are any unknown properties shared by all of them. Another line of research that remains open for the future is to study the plausibility of forecasting changes in a payment network through communities' detection analysis.

**Acknowledgments**

The program Cfinder [41] was used to calculate some network statistics and images. We thank Swedbank AS for allowing us to study this data set. This research was supported by the European Union through the European Regional Development Fund (Project TK 124).